# Predicting Alzheimer's Disease Using 3DMgNet

Yelu Gao[1,2], Huang Huang[2], Lian Zhang[2,*]

[1] School of Mathematics and Computational Science, Xiangtan University, Xiangtan, Hunan, 411105, China

[2] In-Chao Institute, China

*Corresponding author: Lian Zhang (zhanglian@in-chao.com)

## Abstract

Alzheimer's disease (AD) is an irreversible neurode generative disease of the brain. The disease may causes memory loss, difficulty communicating and disorientation. For the diagnosis of Alzheimer's disease, a series of scales are often needed to evaluate the diagnosis clinically, which not only increases the workload of doctors, but also makes the results of diagnosis highly subjective. Therefore, for Alzheimer's disease, imaging means to find early diagnostic markers has become a top priority.

In this paper, we propose a novel 3DMgNet architecture which is a unified framework of multigrid and convolutional neural network to diagnose Alzheimer's disease (AD). The model is trained using an open dataset (ADNI dataset) and then test with a smaller dataset of ours. Finally, the model achieved **92.133%** accuracy for AD vs NC classification and significantly reduced the model parameters.

## 1 Introduction

Alzheimer's disease (AD) is the most common form of geriatric cognitive disorder and accounts for 60% to 70% of cases of dementia. It is a chronic neurodegenerative disease that usually starts slowly and gets worse over time. The most common early symptom is difficulty in remembering recent events (short-term memory loss). As the disease advances, symptoms can include problems with language, disorientation (including easily getting lost), mood swings, loss of motivation, not managing self-care, and behavioral issues [1]. In 2015, there were approximately 46.0 million people worldwide with AD [2]. To date, the cause of AD is still poorly understood and no treatment can stop or reverse its progression has been discovered. Therefore, an earlier identification of the patient with AD related alteration via practical and subjective methods is warranted. It may provide an opportunity for potential disease modified therapy which can cure or prevent the disease in future.

The diagnostic methods of Alzheimer's disease include neuroimaging examination, psychological assessment, and biomarker detection, etc., which play different roles in the early diagnosis of AD, the prediction of the transformation from MCI to AD, and the differentiation of AD from other diseases. Among the above diagnostic methods, neuroimaging technology has been widely used in the field of AD diagnosis. It can provide rich brain structure or brain function information, which is conducive to researchers' analysis of patients' conditions from various aspects. Magetic Resonance Imaging (MRI), functional Magnetic Resonance Imaging (fMRI), Positron Emission Tomography (RET) and Diffusion Tensor Imaging (DTI) are the most common Imaging methods in neuroimaging technology. Due to different imaging techniques, different types of imaging methods have different effects on the diagnosis of diseases. MRI has become the main technical means in the field of AD diagnosis because of its non-invasive, low price and high resolution.

For decades, the role of structural MRI in the diagnosis and evaluation of AD spectrum has been well established. The most robust findings are the atrophy in medial temporal lobe esp. hippocampus structure and can get high performance in patient classification [3]. It is already being accepted as a structural biomarker criterion for AD presence in the prodromal stages [4]. However, a lot of studies have demonstrated that the alteration was not just confined to a few specific brain regions. For example, the posterior cingulate cortex and precuneus, which are both core components of the Default mode network, have been proved involved in early stage of AD [5]. Advances in recently neuroimaging researches also demonstrated cortical atrophy, ventricular enlargement, white matter changes and texture [6] of MRI images can occur in the spectrum of AD and also related to cognitive function alteration. On the other hand, it is difficult to differentiate subtle changes in MRI images between the early stage of AD and normal aging even by experienced clinician. In addition, when performing statistical analysis in traditional structural imaging, it's really hard to combine all of these imaging parameters/variables at the same time, although we all know it's best to incorporate all of this factors in last classification analysis.

Recently, thanks to the development of medical image recognition, some machine learning based methods have been discovered to the diagnosis of AD. Such as SVM, Bayes statistics, voting feature interval classifiers [7] and deep neural networks. In particular, deep neural networks have exceeded all other methods in images classification in recent years because it needs no prior segmentation and feature extraction. The CNN model, as a type of end-to-end architecture, can optimize both the representation of features and the classification performance based on a brain image. There are several studies using deep learning methods for AD, mild cognitive impairment (MCI) and normal controls (NCs) classification [8-12]. Although previous studies based on deep learning model achieve great classification performance for AD diagnosis, 3D convolutional neural network needs to consume more memory space and super high computational cost.

In order to reduce the number of model parameters while maintaining classification performance, we proposed a 3DMgNet model which is a unified framework of multigrid and convolutional neural network. Without complex feature extraction and feature selection processes, our straightforward end-to-end network achieved remarkable classification performance. Our network not only achieved significant classification performance but significantly reduced the model parameters. To test the robustness and generalizability of the imaging biomarkers for AD, we tested it on our own dataset and got excellent classification results.

## 2 Materials and Methods

### 2.1 Data and Preprocessing

The data we have at hand are from the ADNI database(http://adni.loni.usc.edu) and a hospital in China(for convenience, we call it in-house dataset in the rest of this paper).The ADNI dataset has been fully studied around the world. For the in-house dataset, comprehensive clinical details can be found elsewhere in our previous studies [13 - 17]. There are 454 AD subjects and 680 NC in ADNI dataset. There are 75 subjects being AD and 59 subjects being NC in in-house dataset.

To learn valuable information about regional changes in gray matter for the training model, structural MRI images were pre-processed with the standard steps in the CAT12 toolbox (http://dbm.neuro.uni-jena.de/cat/). All structural MRI images were bias-corrected, segmented into gray matter (GM), white matter (WM), and cerebrospinal fluid (CSF) and registered to Montreal Neurological Institute (MNI) space using a sequential linear (affine) transformation. The gray matter images were resliced to 2 mm × 2 mm × 2 mm cubic size, resulting in a volume size of 91 × 109 × 91 with 2 mm$^3$ isotropic voxels.

### 2.2 Network Structure

A simple yet effective unified framework of multigrid and convolutional neural network was proposed for AD diagnosis [18]. We are now in a position to state the main algorithm, namely 3DMgNet as:

---
**Algorithm 1** $u^J = 3DMgNet(f)$

---
1: Input: number of grids $J$, number of smoothing iterations $\nu_\ell$ for $\ell = 1 : J$, number of channels $c_{f,\ell}$ for $f^\ell$ and $c_{u,\ell}$ for $u^{\ell,i}$ on $\ell$-th grid.
2: Initialization: $f^1 = f_{\text{in}}(f)$, $u^{1,0} = 0$
3: for $\ell = 1 : J$ do
4:    for $i = 1 : \nu_\ell$ do
5:       Feature extraction (smoothing):
$$u^{\ell,i} = u^{\ell,i-1} + \sigma \circ B^{\ell,i} * \sigma \left( f^\ell - A^\ell * u^{\ell,i-1} \right) \in R^{c_{u,\ell} \times n_\ell \times m_\ell}. \qquad (1)$$
6:    end for
7:    Note: $u^\ell = u^{\ell,\nu_\ell}$
8:    Interpolation and restriction:
$$u^{\ell+1,0} = \Pi_\ell^{\ell+1} *_2 u^\ell \in R^{c_{u,\ell+1} \times n_{\ell+1} \times m_{\ell+1}}. \qquad (2)$$
$$Sf^{\ell+1} = R_\ell^{\ell+1} *_2 (f^\ell - A^\ell(u^\ell)) + A^{\ell+1} * u^{\ell+1,0} \in R^{c_{f,\ell+1} \times n_{\ell+1} \times m_{\ell+1}}. \qquad (3)$$
$$u^{\ell+1,0} = Avg(u^{\ell+1,0})(\ell < J). \qquad (4)$$
9: end for

---

The input is the normalized 3D gray matter density image, so we replaced the convolution in MgNet with 3D convolution and named it 3DMgNet. In Formula (4), Avg represents average-pooling. In the proposed method, we proved that the addition of average-pooling is more suitable for disease classification than the absence of average-pooling. In the output layer, we use the softmax classifier based on cross-entropy loss.

## 2.3. Experiment setup

Since there are multiple detection images of the same subject in the datasets, in order to avoid data leakage, it is necessary to put multiple detection images of the same subject in the same datasets when dividing the datasets. In accordance with the above principles, we used stratified 10-fold cross-validation, where subjects were randomly partitioned into 10 subsets with stratification based on the subject's label. We repeated the experiments 10 times for AD/NC classification, by using 9 out of 10 subsets for training and the remaining one for testing in each cross-validation round. The accuracy, sensitivity, specificity and area under the curve (AUC) of receiver operating characteristic are used to evaluate the performance of the proposed model.

## 3 Experiment and results

### 3.1 Hyper-parameters

The input is the normalized 3D gray matter density image and the output is a probability for each individual obtained by a soft-max classifier trained with cross-entropy loss. The proposed network was optimized using the SGD algorithm with an initial learning rate of $10^{-4}$, and the batch size was set as 2. We made the following choice of hyperparameters for the 3DMgNet:

J : the number of grids, we choose J = 5 to be consistent with ResNet.

$v_l$ : the number of smoothings in each grids. To be consistent with ResNet-18 we choose $v_l = 2$.

$c_u$ and $c_f$ : the number of feature and data channels, we choose $c_u = 128$, $c_f = 128$.

### 3.2 Comparison with or without average pooling functions

Inspired by a 3D attention network [12], we demonstrated the effects of average pooling functions on classification performance. As shown in Table 1, when average pooling is added to the model, its classification performance is significantly improved.

Table 1. Classification performance with or without average pooling functions.

|  | Accuracy | AUC | Sensitivity | Specificity |
|---|---|---|---|---|
| No average pooling | 0.9074 | 0.9437 | 0.8703 | 0.9365 |
| Average pooling | 0.9213 | 0.9443 | 0.8842 | 0.9500 |

### 3.3 Comparisons with related studies

In the classification of AD and NC, Table 2 presents related studies that conduct classification tasks using deep learning methods based on ADNI dataset.

Table 2. Performance comparison of the proposed method and reported studies on AD classification.

| Methods | Accuracy | AUC | Sensitivity | Specificity | Params |
|---|---|---|---|---|---|
| 3DResNet | 0.9107 | 0.9430 | 0.8656 | 0.9442 | 8,288,290 |
| Jin et al. [12] | 0.9212 | 0.9443 | 0.8842 | 0.9500 | 8,398,882 |
| Proposed | 0.9213 | 0.9443 | 0.8842 | 0.9500 | 6,202,754 |

As shown in Table 2, we reproduced three networks, 3DResNet, 3DAN and 3D MgNet, on the ADNI dataset. The classification performance of our proposed network is much higher than that of 3DResNet, slightly higher than that of 3DAN. In terms of model parameters, compared with the two networks mentioned above, the network that we proposed shows evidently advantages.

**3.4 In-house dataset classification result**

We used the optimal model to test the in-house dataset in 3DResNet, 3DAN and 3DMgNet respectively. The in-house dataset includes 134 MRI data. We use our trained model to test on the dataset directly without fine-tuning the model, and the test results are shown in Table 3.

Table 3. The performance of the proposed method is compared with the existing AD classification studies on in-house dataset.

| Methods | Accuracy | AUC | Sensitivity | Specificity |
|---|---|---|---|---|
| 3DResNet | 86.194 | 0.9561 | 0.7720 | 0.9763 |
| Jin et al. [12] | 83.582 | 0.9568 | 0.7200 | 0.9831 |
| Proposed | 87.911 | 0.9574 | 0.7973 | 0.9831 |

In total, the proposed method generally outperformed the 3DResNet and 3DAN methods in the AD versus NC classifications for the in-house database(Table 3). It is worth noting that the subjects in the in-house dataset and the ADNI dataset came from different countries with different hospitals and different scanners. Although the differences between the data sources was great, the proposed method still achieved a classification accuracy of 87.911%. This indicates that the proposed method has decent generalizability and robustness in dealing with independent datasets, factors which make this method very useful for clinical applications.

## 4 Discussion

While the experimental results demonstrated promising potential clinical applications of the proposed method for AD studies, our study has some limitations that need to be considered in the future. First, the proposed method achieved great results in detecting important regions and diagnosing AD on two databases. However, the performance and robustness of the proposed method should be further validated on a larger population before any actual clinical use. Second, the model is trained based on ADNI dataset, which consists mainly on MRI scan from Western patients. Since there are intrinsic differences between brains from different sides of the world, the model's predicting accuracy on oriental brains, such as Asian brains still needs more dataset to improve. Third, only structural MRI data were used in classification. Since different modalities can provide complementary information for disease diagnosis, we should combine different modality information in the future such as resting-state functional MRI data, diffusion MRI and clinical variables to get more comprehensive understanding for the subjects of early AD stage. Finally, many potential influencing factors have not been taken into account into the model, such as the patient's gender, race, and even genetic variability. These factors are likely to have an impact on the patient's cognitive abilities, and can be further used as input parameters to the model to explore the correlation between these factors and the patient's cognitive abilities. This is also of great significance in medical research. Finally, the model currently only supports the analysis and diagnosis of MRI images. We should also consider the support of multimodal images, such as PiB-PET etc.

## 5 Conclusion

In this research, we proposed a novel 3DMgNet architecture which is a unified framework of multigrid and convolutional neural network to classify AD patients from NC subjects. The proposed method can significantly reduce the model parameters with high classification accuracy. In our own dataset test, the proposed method can also achieve an excellent classification accuracy. The good performance achieved implies that the state-of-the-art methods is able to extract distinct features from the original MRI data that are capable to represent the differences among the early AD patients and normal aging. We believe that after training and validation in larger datasets and multiple centers, this method can be useful to help physicians get more confident diagnosis when facing AD patients in the future.